\documentclass[12pt]{article}
\usepackage[brazilian]{babel}
\usepackage[utf8]{inputenc}
\usepackage[T1]{fontenc}
\usepackage{graphics,graphicx,xcolor}
\usepackage{subfig}                            
\usepackage{longtable}    
\usepackage{url}      
\usepackage{bm}
\usepackage{amsmath, amssymb, amsthm, amsfonts}
\usepackage[hmarginratio=6:6,top=30mm,columnsep=12mm,textwidth=170mm]{geometry}
\usepackage{authblk}
\usepackage{multicol} 
\usepackage{booktabs} 
\usepackage{float} 
\usepackage{hyperref} 
\usepackage{paralist} 
\usepackage{titlesec} 
\newcommand{\beq}{\begin{equation}}
\newcommand{\eeq}{\end{equation}}
\newcommand{\bea}{\begin{eqnarray}}
\newcommand{\eea}{\end{eqnarray}}

\title{\vspace{-10mm}\fontsize{14pt}{11pt}\selectfont \textbf{Osciladores Acoplados, Transferência de Frequências e o Ensino do Mecanismo de Higgs.\\
\vspace{0.5cm}
{\normalsize Coupled Oscillators, Frequency Transfer and the Higgs Mechanism's Teaching}}} % Article title
\vspace{-3mm}
\author{
\normalsize {\bf M. O. Tahim}${}^a$\thanks{Email:makarius.tahim@uece.br},\, \bf C. R. Muniz${}^b$\thanks{Email:celio.muniz@uece.br},\, \bf M. S. Cunha${}^c$\thanks{Email:marcony.cunha@uece.br},\, \bf R. I. O. Júnior${}^a$\thanks{Email:raimundo.ivan@uece.br}\\
 \small ${}^a$Universidade Estadual do Ceará, Faculdade de Educa\c c\~ao, Ci\^encias e Letras do Sert\~ao Central\\ \small CEP 63900-000, Quixadá, CE, Brasil.\\
\small ${}^b$Universidade Estadual do Ceará, Faculdade de Educação, Ciências e Letras de Iguatu\\ \small Av. Dário Rabelo, s/n, Bairro Santo Antônio, CEP 63502-253, Iguatu, CE, Brasil.\\
${}^c$\small Universidade Estadual do Ceará, Centro de Ciência e Tecnologia\\ \small Av. Dr. Silas Munguba, 1700, CEP 60914-903, Fortaleza - CE, Brasil.\\
 }
\date{}
\begin{document}
\maketitle
%\centerline{\rule{8cm}{0.2mm}}
%\thispagestyle{fancy} % All pages have headers and footers
%
\noindent{\footnotesize{\textbf{Resumo:} Neste trabalho, propomos um modelo simples para se ensinar o mecanismo de Higgs de quebra espontânea de simetria em Física de Partículas. A ideia básica está associada com a análise do movimento de duas partículas que, sujeitas a potenciais específicos em determinadas situações, irão reproduzir os movimentos de osciladores harmônicos com o efeito de transferência de frequências.}}\\
{\footnotesize{\textbf{Palavras-chave:} Mecanismo de Higgs, quebra espontânea de simetria, Ensino de Física.}}
 \\[0.5cm]
\noindent{\footnotesize{\textbf{Abstract:} In this work we propose a simple model to teach the Higgs mechanism of spontaneous symmetry breaking in particle physics. The basic idea is associated with the analysis of two particle's motion subjected  to specific potentials that, in certain situations will reproduce oscillatory motions with the frequency transfer effect.}}\\
{\footnotesize{\textbf{Key-words:} Higgs mechanism, spontaneous symmetry breaking, Physics teaching. }}
% \\[0.5cm]
%
\section{Introdução}

\noindent\hspace{1.0cm}{A ideia de ensinar aspectos de física contemporânea acontece em uma grande variedade de espaços: há livros de divulgação científica \cite{abdalla2016}, artigos onde se discutem as ideias mais fundamentais \cite{fis_mod,sasseron,moreira,paulo,araujo}, simulações computacionais \cite{sim_comp,adeil, loop,fosforo}, aplicativos em tablets e smartphones \cite{adeil}, jogos eletrônicos \cite{spracegame} e de tabuleiro \cite{tabuleiro,tabuleiro2}, filmes \cite{interestelar,teor_tudo,vento}, peças teatrais \cite{palco,einstein,radio}, vídeos no YouTube \cite{higgsyoutube,bosonH,discreto_charme} \textit{etc}. Mais recentemente, com o advento do trabalho do Mestrado Nacional Profissional em Ensino de Física (MNPEF)} \cite{mnpef}, fixou-se como uma das metas principais envidar esforços no sentido de tornar realidade o ensino de aspectos de física contemporânea nas escolas de ensino médio. De fato, em suas linhas de pesquisa há a menção clara acerca deste caminho, mais especificamente na \textbf{linha $2$ - Física no Ensino Médio}, onde busca-se a \texttt{"}atualização do currículo de Física para o Ensino Médio de modo a contemplar resultados e teorias da Física Contemporânea visando uma compreensão adequada das mudanças que esses conhecimentos provocaram e irão provocar na vida dos cidadãos\texttt{"} \cite{linhas}.

\noindent\hspace{1.0cm}Do ponto de vista da BNCC  (Base Nacional Comum Curricular), há a determinação de que o ensino médio deve explorar a Física de forma crítica, dando ênfase a temas tais como a exploração do universo e novas tecnologias, temas estes que estão diretamente ligados a assuntos relacionados à Física Moderna e Contemporânea \cite{BNCC}. 
Contudo, como sabemos, a Física é uma área do conhecimento em que a matemática é, na maioria das vezes, um elemento-chave no sucesso de suas previsões e aplicações. Portanto, apesar das várias formas de divulgação, todas elas carecem de uma fundamentação matemática um pouco mais adequada. 

\noindent\hspace{1.0cm}Há uma enorme distância entre uma ideia proposta e o formalismo matemático correspondente.  Um exemplo bastante forte, que agrega no sentido desta aproximação, é o livro \textit{Alice no País da Relatividade: teoria da relatividade para o ensino médio} \cite{geova}, material que entrega Física Moderna com ferramentas matemáticas de nível adequado e apoiadoras da construção conceitual correta. Com isso, queremos dizer que a construção de um modelo matemático, na maioria das vezes, pode se tornar crucial para se ensinar novas teorias físicas, ou seja, não é suficiente apenas termos ideias - é necessário realizá-las. 

\noindent\hspace{1.0cm}Em se tratando de Física Contemporânea, com formalismos matemáticos por demais avançados, o que requer longo treinamento durante anos, é praticamente inviável o ensino para alunos de Ensino Médio e mesmo para alunos de graduação em Física, tanto em licenciaturas como em bacharelados, apesar de tentativas bem intencionadas nesse sentido \cite{Possa,Pimenta}. Recentemente, um belo trabalho preenchendo esta lacuna no  ensino de Física surgiu para a aprendizagem do Mecanismo de Higgs \cite{Helayel}. Nele, os autores utilizam osciladores harmônicos e anarmônicos como modelos de estudo do mecanismo de quebra de simetria e geração de massa para partículas no universo. Também é apresentada uma simulação no software Modellus para implementação de uma prática em sala de aula.
 
\noindent\hspace{1.0cm}O oscilador harmônico é um dos sistemas mais simples que estudamos em Física. Porém, apesar de sua simplicidade, ele possui grandes aplicações em quase todas as áreas da Física, de baixas a altas energias. Esse sistema é, por exemplo, a base para se entender como aplicar mecanismos de quantização aos campos de toda sorte, ou seja, ele nos ensina a entender as partículas como oscilações localizadas dos campos \cite{Marcelo Gomes}. Como disse certa vez Sidney Colemann (1937-2007): \texttt{"}A carreira de um jovem físico teórico consiste em tratar o oscilador harmônico em níveis crescentes de abstração\texttt{"} \cite{Colemann}. Em particular o sistema físico que nos interessa aqui é justamente o de dois osciladores acoplados. A interação entre eles produz um movimento bastante interessante: em certas condições, a energia dos osciladores é trocada de um para o outro, fazendo com que um deles oscile enquanto o outro não \cite{acoplado,coupled_osc,Symon}. Este movimento, em especial, lembra o mecanismo físico produzido pela partícula de Higgs quando na transição de fase e correspondente geração de massa: em certo momento temos duas partículas, sendo uma massiva (partícula de Higgs) e outra não-massiva; depois da transição de fase, estas características são, em simples termos, trocadas entre elas. É neste caso, exceto se a partícula for colocada em repouso na origem, ela se afastará indefinidamente. Caso ela seja colocada em repouso na origem, qualquer perturbação feita sobre ela, por menor que seja, fará com que ela se afaste indefinidamente(ou seja, a partícula experimenta uma força repulsiva que a empurra para longe)ste aspecto que queremos explorar.

\noindent\hspace{1.0cm}Neste sentido, propomos neste trabalho complementar a discussão feita em \cite{Helayel} acrescentando efetivamente dois aspectos: $(1)$ a interação de dois osciladores de modo a demonstrar o análogo do mecanismo real de geração de massa e $(2)$ uma modelagem para a mudança de sinal da constante de mola para justificar o potencial de quebra de simetria para baixas energias. A matemática envolvida é ainda bastante simples e precisa-se apenas que os alunos tenham conhecimentos associados com a física de osciladores harmônicos. Neste modelo apresentamos um \texttt{"}oscilador de Higgs\texttt{"} e apontamos a frequência $\omega$ do oscilador harmônico como o análogo da massa de uma partícula \cite{Ivan}.

\noindent\hspace{1.0cm}O trabalho é organizado da seguinte maneira. Na seção $2$ relembramos a física de osciladores discutindo o sinal da constante de mola de modo a montar a ideia de quebra de simetria. Na seção $3$ discutimos o modelo com dois osciladores e apresentamos o análogo do mecanismo de quebra de simetria e geração de massa. Na Seção $4$ fazemos um breve relato das ideias do Mecanismo de Higgs. Na última seção, complementamos este trabalho apresentando uma simulação computacional para aplicação prática em sala de aula. Por fim, apresentamos conclusões e considerações finais.

\section{O Oscilador Harmônico}

\noindent\hspace{1.0cm}Nesta seção tomaremos em consideração o modo de trabalho realizado em \cite{Helayel}, ou seja, variações das funções entram no lugar do conceito de derivadas. Para iniciarmos, lembremos do sistema massa-mola, mais especificamente que sua energia total (desconsiderando o atrito entre o bloco e o apoio) é dada por
\begin{equation}\label{oscilador}
E=\frac{1}{2}mv_{h}^{2}+\frac{1}{2}kh^{2},
\end{equation}
onde $m$ é a massa do oscilador, $k$ a constante da mola, $v_{h}$ a velocidade do oscilador e $h$ a deformação da mola (denotaremos a posição pela letra \textit{h} pois logo mais faremos considerações sobre a partícula de Higgs). Há aqui algumas observações importantes que serão úteis para nossa discussão posterior. A primeira delas é quanto à análise da energia potencial elástica. Como o sinal da constante $k$ é positivo, a energia potencial é descrita por uma parábola com mínimo em $h=0$ e concavidade para cima. O mínimo é estável e temos então a oscilação usual (por causa de forças restauradoras) em torno desse mínimo, com amplitude dependente da energia total do oscilador.

\noindent\hspace{1.0cm}De fato, temos uma oscilação que classificamos como harmônica, pois usamos funções harmônicas (senos e cossenos) como soluções para tratar o movimento da forma mais completa possível. A outra observação importante é que este sistema possui uma simetria sob a transformação $h\rightarrow -h$, significando simplesmente que a energia não muda com essa operação (podemos dizer igualmente que a energia é invariante sob aquela transformação). Esta simetria é chamada de paridade $P$ (ou de reflexão especular). Em outras palavras, trocar $h$ por $ -h$ inverte o sentido do movimento (ou seja, troca o sinal da velocidade) e como a velocidade aparece ao quadrado na expressão da energia, o termo de energia cinética não muda sob paridade. Também, o termo de potencial é uma função par, característica essa fundamental para manutenção da simetria citada e que concorre para gerar o gráfico da energia potencial simétrico em relação ao eixo vertical.

\noindent\hspace{1.0cm}Outra simetria que podemos observar é a que ocorre quando trocamos $t$ por $-t$. Havendo ou não troca de sinal na definição de $h$ ao fazermos essa operação, os termos quadráticos indicam que a energia fica invariante sob tal transformação, designada por {\it inversão temporal}. Ou seja, se assistirmos a um filme do sistema em funcionamento, não teríamos condições de averiguar se a película está rodando de trás para a frente ou ao contrário. Em outras palavras, perderíamos a noção de {\it flecha do tempo}, que é o sentido do fluxo do tempo sempre ir do passado para o futuro.

\noindent\hspace{1.0cm}Uma maneira de se aprofundar mais na física deste sistema pode ser feita simplesmente trocando-se o sinal da constante $k$ na Eq. \eqref{oscilador}, que descreve a energia total do oscilador, especificamente em sua energia potencial, de positivo para negativo. Nesta nova situação, a energia potencial continua sendo representada por uma parábola, no entanto agora com concavidade para baixo, ou seja, com um máximo local em $h=0$. Neste caso, a partícula que representa o oscilador se afastará indefinidamente. A exceção é se ela for colocada em repouso na origem, onde $E=0$, e assim não haverá movimento e a partícula restará parada em equilíbrio. Entretanto, esse equilíbrio é instável e qualquer perturbação feita sobre ela, por menor que seja, fará com que se afaste também indefinidamente (ou seja, a partícula experimentará uma força repulsiva que a empurrará para longe). Em outras palavras, não temos uma oscilação aqui. E quanto à simetria de paridade?  O fato é que ela ainda existe neste caso sem, no entanto, descrever o movimento de um oscilador harmônico. Veremos, logo adiante, que podemos construir um modelo para o qual se a partícula se deslocar para longe da origem fará com que esta simetria desapareça.

\noindent\hspace{1.0cm}Em resumo, apresentamos duas situações distintas. Em uma delas temos uma partícula que possui oscilação harmônica e que possui a simetria de paridade; na outra, não há oscilação, apesar de termos a referida simetria. A pergunta que pode ser feita a esta altura é a seguinte: é possível construir de uma só vez um modelo que contemple estas duas situações? A resposta é {\it sim} e passaremos a discutí-lo agora. Neste sentido, vamos propor agora um oscilador com um termo de anarmonicidade cuja energia total é dada por
\begin{equation}\label{h}
E=\frac{1}{2}m v_{h}^{2}+\frac{1}{2}k(T)h^{2}+\lambda \frac{h^{4}}{4}.
\end{equation}
\noindent A novidade é que temos agora uma constante da mola $k(T)$ que depende explicitamente de um parâmetro físico, que aqui consideramos ser a temperatura $T$, e uma interação de quarta ordem em $h$, o termo de anarmonicidade. A dependência da constante com a temperatura é usual se quisermos discutir sistemas onde temos transições de fase. Todos sabemos que a água, se mudarmos sua temperatura, passa por várias fases. Há inúmeros exemplos onde se pode discutir este aspecto. O importante é que a mudança da temperatura produz mudança de fase, ou seja, uma quebra de simetria do sistema estudado. No caso da água, quando em estado líquido, possui simetria de rotação e, quando baixamos a temperatura até o ponto de congelamento, há quebra de simetria quando da formação dos cristais de gelo (o sistema define uma direção na formação do reticulado cristalino). No caso de nosso oscilador, a constante $k$ determina as características físicas da mola, ou seja, nada mais natural do que ela possuir certa dependência com a temperatura.

\noindent\hspace{1.0cm}{Quanto ao termo de ordem quártica, podemos considerá-lo como um termo que gera não-linearidades, significando que as soluções não obedecem ao princípio de superposição válido para o caso simplesmente harmônico, isto é, sem a presença daquele termo (a lei de Hooke não é mais válida)\cite{Filipponi}. 
}
\begin{figure}[ht]
\center
\includegraphics[width=10cm,height=8cm]{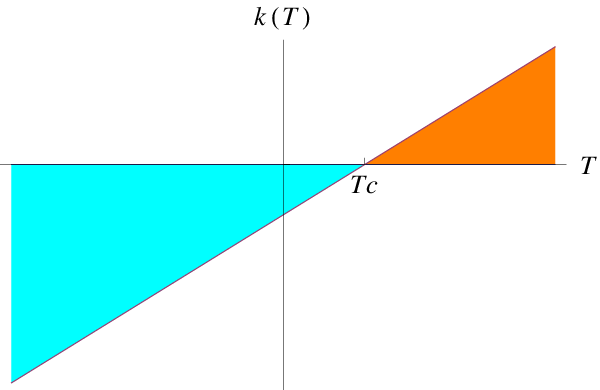}\hspace{1cm}
\caption{Modelo para a varia\c{c}\~ao da constante da mola com a temperatura, $k(T)=\alpha T-\beta$. Chamamos $T_c$ a temperatura crítica, aquela na qual $k(T)$ é nula e marca a mudança de seu sinal.} \label{figk} 
\end{figure}

\noindent\hspace{1.0cm}Vamos propor agora um modelo de variação da constante $k$ com a temperatura. A ideia é traçar uma analogia com as mudanças de temperatura do Universo, que nasceu muito quente e diminuiu sua temperatura com o passar do tempo \cite{Ronaldo}. Para nosso objetivo, é suficiente considerar a forma mais simples $k(T)=\alpha T - \beta$, com $\alpha$ e $\beta$ números reais positivos ($\alpha$ tem dimensão de $N/mK$ e $\beta$ dimensão de $N/m$, de modo a termos a dimensão correta para a constante de mola - adotamos o sistema MKS de unidades de medida). Desta feita, é possível ver que os valores de $k$ variam de negativo a positivo, sendo zero em $T_C={\beta}/{\alpha}$. Ainda mais, em altas temperaturas o valor de $k$ é positivo, enquanto para baixas temperaturas, seu valor é negativo [Fig. \eqref{figk}]. É importante agora discutir como se comporta o gráfico do novo potencial presente na Eq. \eqref{h}, ou seja,
\begin{equation} \label{potencial}
V(h)=\frac{1}{2}k(T)h^{2}+\lambda \frac{h^{4}}{4}.
\end{equation}

\begin{figure}[ht] 
\center
\includegraphics[width=10cm,height=8cm]{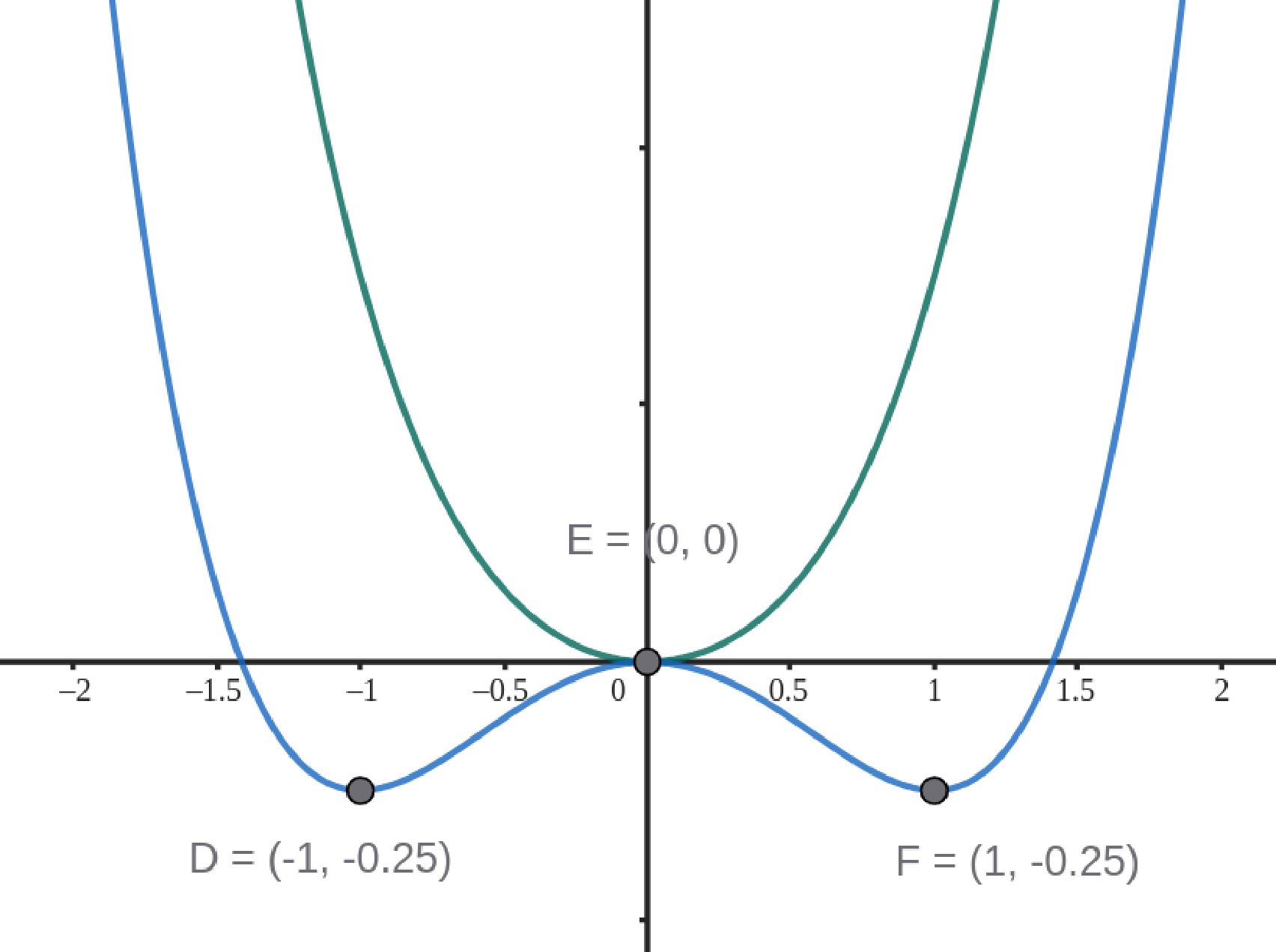}
\caption{Gráficos do novo potencial [Eq. \eqref{potencial}]. A curva na cor verde possui um mínimo quando $k(T)=1$ e $\lambda =1$. A curva em azul possui dois mínimos quando $k(T)=-1$ e $\lambda=1$. Ao escolhermos os valores de $k(T)$ e $\lambda$, os mínimos, na segunda situação, são $h=-1$ e $h=1$.} \label{minimos}
\end{figure}

\noindent\hspace{1.0cm}É fácil verificar que quando $T>{\beta}/{\alpha}$ temos então $k>0$ e o gráfico apresenta apenas um mínimo em $h=0$. Quando $T<{\beta}/{\alpha}$ e $k<0$, teremos desta vez um máximo local em $h=0$ e mínimos locais em $h=\pm a$, onde $a=({-k}/{\lambda})^{\frac{1}{2}}$ [Fig. \eqref{minimos}]. Basta agora discutir os movimentos desse oscilador em cada caso. Na primeira situação, em altas temperaturas, o sistema deve permanecer oscilando em torno do mínimo em $h=0$. Podemos considerar pequenas oscilações e, neste caso, teremos o conhecido oscilador harmônico de volta, pois o termo quártico da energia potencial é desprezível em relação ao termo quadrático. Importante ressaltar que a simetria de paridade $P$ permanece intacta a estas temperaturas. A surpresa acontece quando vemos a situação física à medida que a temperatura diminui, de modo a termos $k<0$. Nesta situação, o sistema tem que escolher em qual dos dois mínimos de energia ele ficará, já que na posição do máximo local o sistema é instável. O sistema possui uma característica, a constante da mola, que depende da temperatura. De outra maneira, o sistema não é fechado pois a mudança de temperatura se dará por fatores externos a ele, e a energia se conserva, caso consideremos também o meio externo. Mais especificamente, se considerarmos uma energia inicial para o oscilador na situação $k>0$ teremos um mínimo de energia com valor mais alto que os mínimos correspondentes para $k<0$: o sistema definitivamente terá de ir para um deles! Também, por conta disso, a energia total do oscilador mudará: há calor sendo trocado entre o sistema e o meio externo. 

\noindent\hspace{1.0cm}O processo de transição de fase depende criticamente de um mecanismo externo de termalização, responsável por conduzir o sistema a um novo estado físico. Nesse contexto, a dinâmica é governada pelo sinal da constante elástica efetiva, cujo valor é termicamente controlado — resultando em uma modificação significativa do perfil do potencial. Consequentemente, a energia da partícula é reajustada, levando-a a oscilar em torno de um dos novos mínimos de potencial estabelecidos após a transição. Na Fig. \eqref{minimos} abaixo, o gráfico em verde representa uma situação para altas temperaturas, onde o sistema oscila em torno da origem. Para a situação em que a temperatura diminui, temos o gráfico em azul, em que o sistema terá que escolher um dos mínimos (pontos $D$ e $F$ no gráfico).

\noindent\hspace{1.0cm}Supondo que o sistema escolheu $h=+a$, devemos avaliar, como usual, a situação física nas vizinhanças desse mínimo. Para tal, fazemos uma mudança de coordenadas $h\rightarrow h'=h-a$ {na eq. (\ref{h}), uma simples mudança na origem},  e avaliamos a energia do sistema neste caso. O resultado (fazendo $k=-k'$, onde $k'>0$) é dado por

\begin{equation}\label{Mk3}
E={\frac{1}{2}mv_{h'}^{2}}+a(\lambda a^{2}-k')h'+\frac{1}{2}(3\lambda a^{2}-k')h'^{2}+\lambda ah'^{3}+\lambda \frac{h'^{4}}{4}.
\end{equation}

\noindent\hspace{1.0cm}Na energia acima, desconsideramos termos constantes, pois estes não contribuem para as equações do movimento ($v_{h'}$ é a velocidade nas proximidades do mínimo). O que aconteceu com a simetria de paridade $P$? Considerando que $(\lambda a^{2}-k')h'=0$ para $a=({-k}/{\lambda})^{{1}/{2}}$, com $k=-k'$, ela já não mais existe: a nova energia possui agora termos de potência ímpar, pois, quando fazemos $ h\rightarrow h'=h-a$, colocamos a origem sobre o ponto de mínimo em $h = a$. Com essa nova origem, a energia potencial deixa de ser uma função par. Considerando pequenos deslocamentos, recuperamos uma partícula com movimento de oscilação como demonstrado em \cite{Helayel}. Esse processo, em Física de Partículas, é denominado de quebra espontânea de simetria, ou seja, os mínimos são equivalentes mas o sistema somente pode ocupar um desses estados. 

\noindent\hspace{1.0cm}Em resumo, temos duas situações que podem ser atingidas por este sistema: a primeira em altas temperaturas, se consideramos pequenas oscilações, temos de fato um oscilador harmônico que obedece à simetria de paridade; a segunda situação se dá em temperaturas mais baixas, onde se perde a simetria citada, mas ainda recuperando oscilação harmônica para pequenas oscilações.  Podemos afirmar que a temperatura crítica $T={\beta}/{\alpha}$, que define o limite entre essas duas situações, corresponde, portanto, à temperatura de transição de fase do sistema.

\noindent\hspace{1.0cm}Como já afirmamos, a ideia deste trabalho é tomar a analogia acima como partida para ensinar de maneira mais simples o mecanismo de Higgs. Neste caso, podemos interpretar a situação descrita como a de uma partícula massiva em altas temperaturas (numa situação de simetria) e, quando da diminuição da temperatura, a simetria é \texttt{"}quebrada\texttt{"} e a partícula perde sua massa. A questão é que, no mecanismo de Higgs usual, há pelo menos duas partículas em jogo, uma massiva (o campo de Higgs) e outra não massiva, mas que interage com o campo de Higgs. Até o momento discutimos um modelo de oscilador que contém apenas uma partícula. Discutiremos, na próxima seção, como um modelo de dois osciladores interagentes entre si pode nos ensinar as ideias mais básicas a respeito do mecanismo de Higgs.

\section{Quebrando Simetrias}

\noindent\hspace{1.0cm}Nesta seção, incluiremos uma segunda partícula à nossa discussão (sua coordenada, agora sim, denotaremos por $x(t)$). Seu movimento dependerá, como veremos, de como ela interage com a partícula $h(t)$. Para simplificar o raciocínio, denominaremos as partículas de $A$ e $h$, sendo esta última a que denominamos ``oscilador de Higgs''. Propomos a seguinte energia total deste sistema:
\begin{equation}\label{broken}
E(x,h)=E(x)+I(x,h)+E(h)=\frac{1}{2}v_{A}^{2}+\frac{1}{2}\Lambda h^{2}x^{2}+ \frac{1}{2}v_{h}^{2}+\frac{1}{2}\omega^2(T)h^{2}+\lambda \frac{h^{4}}{4},
\end{equation}
onde redefinimos velocidades e os deslocamentos $\sqrt{m_h}v_h\rightarrow v_h$, $\sqrt{m_A}v_A\rightarrow v_A$, $\sqrt{m_h}h(t)\rightarrow h(t)$ e $\sqrt{m_A}x(t)\rightarrow x(t)$ de modo a absorver as massas newtonianas $m_h$ e $m_x$ das partículas em jogo e  reinterpretar $\omega(T)$ como a \texttt{"}verdadeira\texttt{"} massa em nosso propósito de discutir o mecanismo de Higgs. Também denotamos as velocidades específicas de cada oscilador, ou seja, $v_A$ e $v_h$. É simples ver as redefinições acima pois, por exemplo, podemos escrever 

\begin{equation}
\frac{1}{2}m_{h}v^{2}_{h}=\frac{1}{2}\left(\sqrt{m_h}v_h\right)^2\rightarrow\frac{1}{2}v_{h}^2. 
\end{equation}

\noindent Para o termo contendo a constante da mola escrevemos 
\begin{equation}
 \frac{1}{2}k(T)h^2=\frac{1}{2}\frac{k(T)}{m_{h}}m_{h}h^2=\frac{1}{2}\omega^{2}(T)(\sqrt{m_h}h)^2\rightarrow \frac{1}{2}\omega^{2}(T)h^{2}.   
\end{equation}

\noindent\hspace{1.0cm}Na última equação acima, $\omega^{2}(T)=k(T)/{m_{h}}$. Dessa vez, a frequência ou massa é que depende da temperatura $T$. Vemos claramente duas partes de energia independentes uma da outra, $E(x)$ e $E(h)$, sendo esta última base da discussão da seção anterior. A parte da energia denotada por $I(x,h)=\frac{1}{2}\Lambda h^{2}x^{2}$ é o que se chama termo de interação entre as partículas $A$ e $h$: sem este termo as duas partículas não influenciariam uma à outra. A quantidade $\Lambda$ se chama constante de acoplamento e mede o quão forte é a interação entre as partículas (ou seja, faz o papel, por exemplo, da carga elétrica ou da constante gravitacional). Vale ressaltar que o termo de interação envolve o produto do quadrado dos deslocamentos $x$ e $h$. Uma interação deste tipo surge em problemas de dois osciladores acoplados considerando anarmonicidades também \cite{Kuros:2010dxu,Banks:1973ps}.

\noindent\hspace{1.0cm}A simetria deste modelo parecerá bastante simples se notarmos que, ao fazermos $h\rightarrow -h$, a energia não muda, como antes discutido. O mesmo acontece se fizermos $x\rightarrow -x$, ou seja,  a energia total novamente não muda, mesmo considerando o termo $I(x,h)$ de interação: todos os termos da energia total tem potências pares em $x$ e $h$. Estas duas simetrias são de paridade $P$ e, pelo fato de elas estarem associadas a partículas diferentes, denotamos que a simetria total deste sistema é $P\otimes P$. O símbolo $\otimes$ nos diz que estas simetrias são efetivamente misturadas pelo termo de interação $I(x,h)$. Acaso este termo não existisse, isto é, não houvesse interação ($I(x,h)=0$), ainda assim teríamos as duas simetrias, no entanto agora de uma maneira mais trivial a qual denotaríamos como uma soma simples $P\oplus P$.

\noindent\hspace{1.0cm}Dadas todas as características deste modelo, cabe agora a seguinte pergunta: o que acontece quando mudamos a temperatura $T$? Devemos lembrar do comportamento de $k(T)$ (agora $\omega^2(T)$) e descrever o comportamento de ambas as partículas. Neste sentido é o potencial $V(h)$, já discutido, que definirá estes comportamentos. Sendo assim, para a partícula $h$, quando $T>{\beta}/{\alpha}$ então $\omega^2>0$ e o gráfico para $V(h)$ apresenta apenas um mínimo em $h=0$. Neste caso, em altas temperaturas, o sistema deve permanecer oscilando em torno do mínimo em $h=0$. Podemos considerar pequenas oscilações e teremos novamente o oscilador harmônico. 

\noindent\hspace{1.0cm}Para a partícula $A$, a conclusão que rapidamente chegamos é que ela continua sentindo a presença da partícula  $h$, ou seja, seu movimento é determinado pelas pequenas oscilações da partícula $h$. Efetivamente, porém, a partícula $A$ não se comporta como um oscilador harmônico: não há um termo de força restauradora. Na interpretação aqui adotada, essa partícula não possui massa.

\noindent\hspace{1.0cm}Resta agora discutir o que acontece quando diminuímos a temperatura. Quando $T<{\beta}/{\alpha}$, $\omega^2<0$ e teremos agora um máximo local em $h=0$ e mínimos locais em $h=\pm a$. O sistema deve escolher para qual mínimo irá \texttt{"}rolar\texttt{"}. Novamente vamos supor que será $h=+a$. Como já sabemos, devemos avaliar a física nas proximidades desse mínimo e, para tal, fazemos a mudança de referencial para $h\Rightarrow h'=h-a$ e calculamos a energia do sistema. Para a partícula $h$ concluímos exatamente o mesmo que na seção anterior, ou seja, nossa nova partícula $h'$ é um oscilador, mas sua simetria $P$ é perdida. Agora a surpresa: a partícula $A$ se transforma em um oscilador harmônico e sua simetria $P$ permanece intacta! Para ver isso, devemos avaliar a transformação $h\rightarrow h'=h-a$ no termo de interação $I(x,h)$, o que nos fornece
\begin{equation}
I(x,h)=\frac{1}{2}\Lambda h^{2}x^{2}\rightarrow \frac{\Lambda}{2}a^{2}x^{2}+\Lambda ah'x^{2}+\frac{\Lambda}{2}h'^{2}x^{2}.
\end{equation}

\noindent\hspace{1.0cm}Claramente, o termo que depende somente de $x$ é responsável por uma força restauradora. Caso consideremos pequenas oscilações em $x$ e $h$, teremos exatamente a conclusão já aludida. Em resumo, temos o seguinte: quando a simetria é do tipo $P\otimes P$ em altas temperaturas, $h$ é oscilador e $A$ não é oscilador (considerando pequenas vibrações); quando a temperatura baixa além de um limite específico, quebramos a simetria $P\otimes P$ para somente $P$ e agora $A$ é oscilador harmônico  com ``massa'' efetiva $\omega=\sqrt{\Lambda} a$ com $h'$ oscilador também (considerando pequenas vibrações), ou seja, $h'$  atua como partícula \texttt{"}geradora\texttt{"} de nova oscilação. Em outras palavras, é basicamente isso o que acontece no mecanismo de Higgs: massas de partículas são geradas ao custo de simetrias serem quebradas. 

\section{O Mecanismo de Higgs}
\noindent\hspace{1.0cm}O Mecanismo de Higgs é tema bastante importante em Física Contemporânea, mais especificamente em Física de Partículas \cite{Griffiths}. Embora o modelo de geração de massa tenha sido proposto na década de $60$ do século passado, somente em $2012$ a partícula de Higgs foi encontrada no acelerador de partículas Large Hadron Collider (LHC), localizado na fronteira franco-suíça \cite{CMS}. 

\noindent\hspace{1.0cm}A matemática envolvida está associada com teorias de grupos de simetrias em Física de Partículas, além dos formalismos usuais de teorias de campos. Em resumo, a partícula de Higgs encontrada completa o Modelo Padrão de Partículas, especificamente no que concerne à montagem do modelo Eletrofraco. Neste modelo a simetria importante é dada pelo grupo denominado $SU(2)\otimes U(1)$ \cite{Ryder} e entende-se que, por conta dela, as forças fraca e forte estão unificadas em escalas de altas energias (ou altas temperaturas, em um contexto cosmológico).

\noindent\hspace{1.0cm}Nesta primeira situação, de altas energias,  a partícula de Higgs possui massa e todas as demais partículas do modelo são não massivas. Quando a temperatura do Universo diminui, há uma quebra (espontânea) da simetria $SU(2)\otimes U(1)$ para o grupo $U(1)$, com o Higgs perdendo a sua massa e algumas partículas do modelo adquirindo-a (o elétron e as partículas mediadoras da força fraca $W^{\pm}$ e $Z^{0}$). O fato é que partículas mediadoras de força, que possuem massa, têm alcance finito e esse é justamente o efeito que se busca explicar com o Mecanismo de Higgs.

\noindent\hspace{1.0cm}Dada a descrição resumida apresentada acima, nota-se rapidamente o quanto o discurso matemático dificulta o entendimento de uma ideia fundamental. É justamente essa dificuldade de discurso a motivadora de prosseguimento à discussão proposta. 

\section{Uma Proposta de Modelagem para a Sala de Aula.}

\noindent\hspace{1.0cm}Depois de termos discutido os aspectos matemáticos, acreditamos que uma prática envolvendo uma simulação computacional pode complementar a compreensão do mecanismo de Higgs e a quebra de simetria (seguindo as ideias em \cite{Helayel}). Em resumo, podemos discutir tais temas usando gráficos dos potenciais aqui estudados. Para isso, propomos uma atividade utilizando o Geogebra, uma ferramenta de fácil acesso pois pode ser encontrada de forma gratuita na PlayStore e AppleStore, assim como em plataforma online. Dessa forma o professor pode utilizar o \textit{smartphone} como um aliado para suas aulas. 

\noindent\hspace{1.0cm}O Geogebra usa linguagem bastante simples, tanto para o professor quanto para o aluno. As equações são digitadas facilmente através do teclado, e tem-se a liberdade de interagir de forma intuitiva com as possibilidades de visualização de gráficos. Aqui, propomos a prática em três momentos: No primeiro, os alunos são convidados a se familiarizar com o Geogebra. Para isso, apenas são plotadas duas curvas, representando a energia do oscilador simples, sendo que umas das curvas tem concavidade para cima (ponto de mínimo local) e a outra para baixo (ponto de máximo local).
No segundo momento, introduzimos o termo de anarmonicidade e pedimos para que se observe a transição que ocorre no gráfico a medida que a temperatura diminui. O gráfico passa de apenas um mínimo (em altas temperaturas) para um máximo e dois mínimos à medida que baixamos a temperatura.  No terceiro momento discute-se a ideia de quebra de simetria analisando-se a simetria de paridade dos gráficos desenhados. Apresentamos a seguir as propostas de práticas. 
\subsection{Prática 1.}
\begin{figure}[H]
\center
{\includegraphics[width=8cm,height=5cm]{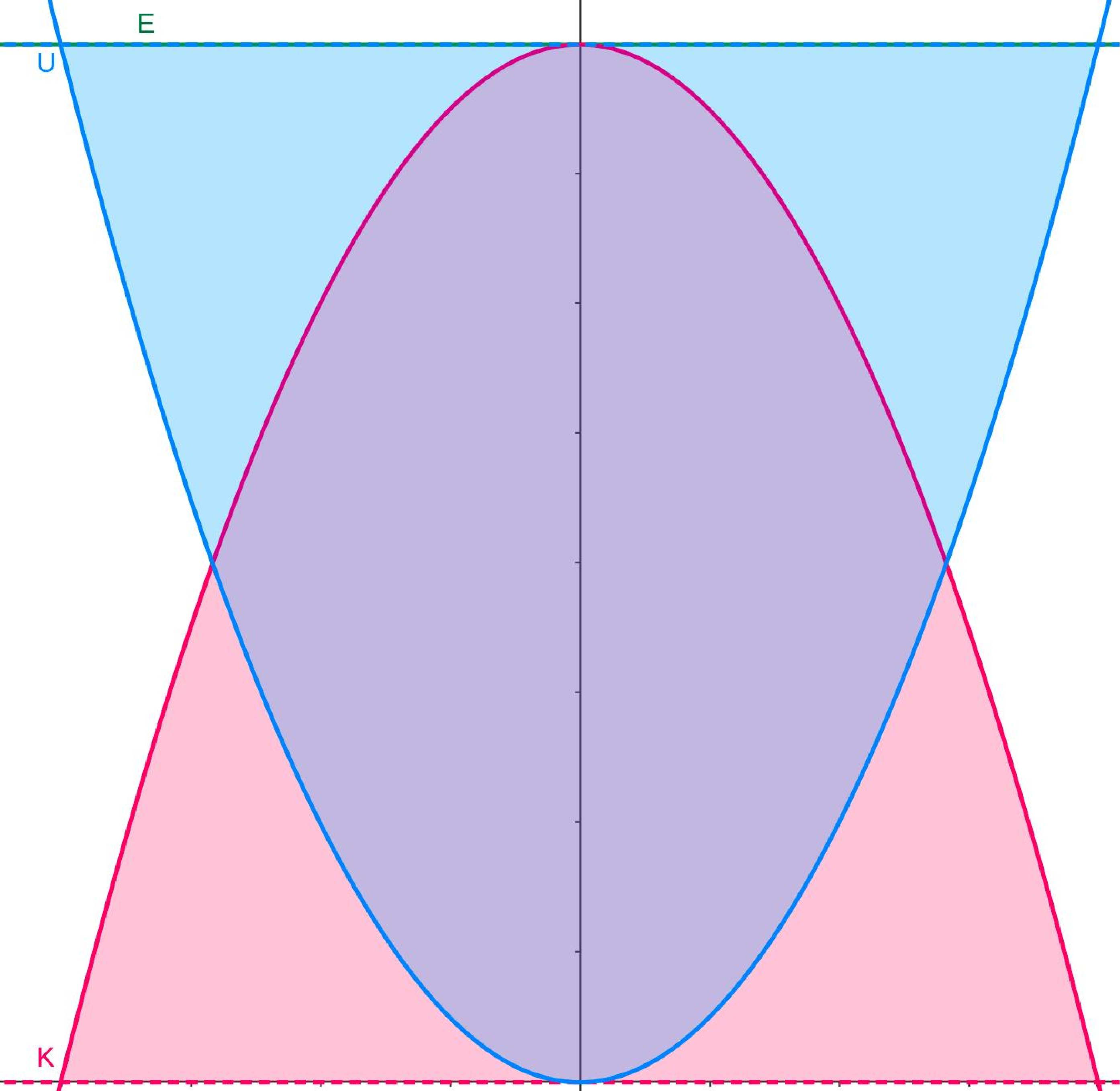}
}%\hspace{1cm}
\caption{\label{fig_pot_2D}Representação bidimensional do potencial do oscilador harmônico simples.}
\end{figure}

\noindent\hspace{1.0cm} Sabemos das aulas de física do ensino médio, que a energia do Oscilador Harmônico é dada pela equação \eqref{oscilador}, e possui o comportamento apresentado na Fig. \eqref{fig_pot_2D}.

\noindent\hspace{1.0cm}Para construirmos essa figura, basta fazermos $k=1$ na equação (\ref{oscilador}). Em seguida, pede-se ao aluno que digite a inequação $ y \geq \frac{1}{2} x^{2}$ (curva em azul com preenchimento também azul, mas que pode ter uma outra cor escolhida pelo aluno). Depois, pede-se que seja digitada a inequação $  0 \leq y \leq - \frac{1}{2}x^{2} + 8 $ (que vai produzir a curva em vermelho com preenchimento também em vermelho). Após isso, constrói-se a reta $y=8$ (basta digitar $y=8$ na caixa de interação do programa e a reta será desenhada, ou escolher qualquer número, mas tomando-se o cuidado com a escala numérica em questão). 
Ao final desse primeiro momento, o aluno deve perceber a forma geral do gráfico do oscilador harmônico (uma parábola, cuja concavidade depende do sinal de $k$)

\subsection{Prática 2.}
\begin{figure}[H]
\center
{\includegraphics[width=18cm,height=8cm]{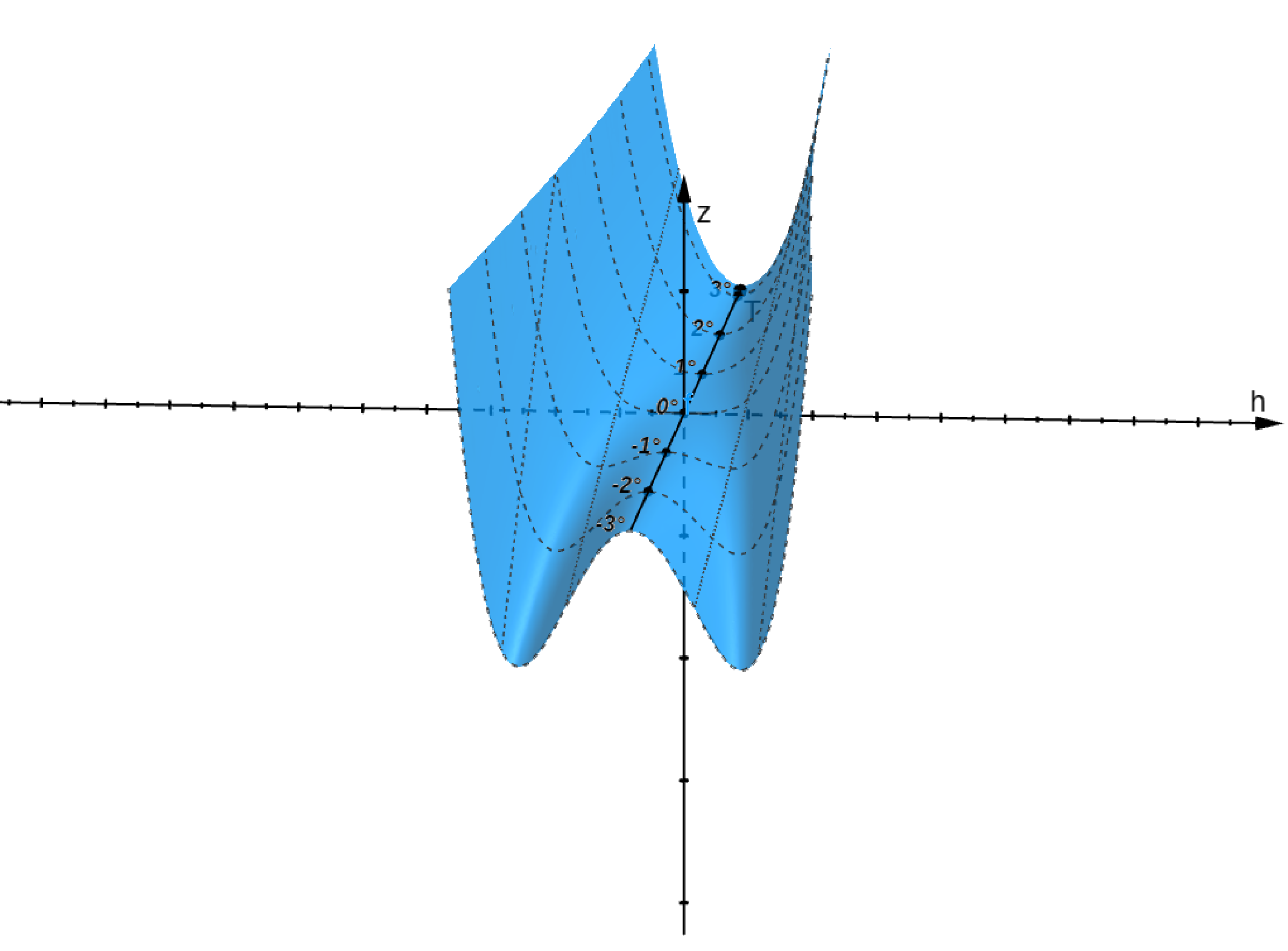}}
\caption{Representação tridimensional do potencial do oscilador anarmônico.} \label{fig_pot_3D} 
\end{figure}

\noindent\hspace{1.0cm}Para a realização dessa prática, precisamos utilizar a ferramenta \textit{Calculadora} $3D$ do Geogebra, uma vez que desejamos ilustrar graficamente a equação \eqref{potencial} e, além disso, a energia agora depende da coordenada $h$ e da temperatura. A  Fig. \eqref{fig_pot_3D} mostra a transição entre valores de $ k>0$ para $ k<0$. Na figura, adotamos $ \alpha=\beta=1$ e $\lambda=2$. 

\noindent\hspace{1.0cm}Para a construção dessa prática procedemos da seguinte forma. Digitamos a seguinte equação no Geogebra $ z= \frac{1}{2} yx^{2} + \frac{1}{2} x^{4} $ (perceba que $x$ faz o papel da coordenada $h$ e $y$, o papel da temperatura). Configuramos o eixo $y$ para ir de $-3$ até $3$; ao visualizarmos o gráfico, vemos que para $ T>1 $ ele possui um mínimo em $h=0$ e, para $T<1$, o potencial passa a ter um máximo em $h=0$ e dois mínimos simétricos.

\subsection{Prática 3.}

\noindent\hspace{1.0cm}Nesta terceira parte, estudamos de fato o significado de quebra de simetria analisando diretamente o comportamento dos gráficos dos potenciais nas situações que apresentamos. Para tal, basta checar a simetria de paridade do gráfico de potencial efetivo para as partículas $A$ e $h$ após a translação para o mínimo, quando $\omega^{2}(T) <0$ na equação (\ref{broken}). 

\begin{figure}[H] 
    
    \label{comparação}
    \subfloat[\label{fig:seu_madruga}Potencial com simetria de paridade com relação ao eixo $y$(Os valores de $x$ estão representados na linha horizontal).]{
    \includegraphics[width=0.48\textwidth]{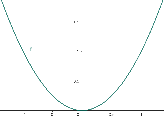}
} \hfill
\subfloat[\label{fig:seu_barriga}  Potencial sem simetria de paridade com relação ao eixo $y$ (Os valores de $h$ estão representados na linha horizontal).]{
    \includegraphics[width=0.39\textwidth]{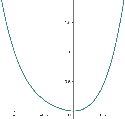}
}
\caption{Gráficos para a prática 3.}
\end{figure}

\noindent\hspace{1.0cm}Para a partícula $A$ temos que, para pequenas oscilações, o seu potencial efetivo pode ser visualizado usando-se a equação $y=x^2$ [Fig. \eqref{fig:seu_madruga}]. Vê-se que o potencial é simétrico em relação ao eixo $y$ revelando a simetria de paridade remanescente após baixarmos a energia do sistema. Neste caso, a partícula $A$ tem um potencial efetivo de Oscilador Harmônico e, segundo a interpretação apresentada, é partícula \texttt{"}massiva\texttt{"}. 

\noindent\hspace{1.0cm}Para a partícula $h$, o seu potencial efetivo (após translação para o mínimo) pode ser descrito pela equação $y=h^2+h^3+h^4$ (aqui colocamos as constantes da Eq. (\ref{Mk3}) todas iguais a $1$). Vemos que o gráfico desta função não apresenta a simetria de paridade anterior [Fig. \eqref{fig:seu_barriga}]. Isso acontece por causa da translação de eixo feita para avaliarmos a física do sistema nas proximidades do mínimo, para pequenas oscilações. Apesar da simetria perdida, vemos que o termo de menor potência ainda é do tipo $h^2$ para pequenas oscilações. Isso significa que a partícula $h$ ainda oscila, sendo responsável pela geração de oscilação da partícula $A$ na situação de interação discutida nas seções anteriores.

\section*{Conclusões}

\noindent\hspace{1.0cm}Neste artigo, utilizando conceitos matemáticos bastante acessíveis,  discutimos um modelo didático para ensinar as ideias básicas do mecanismo de Higgs e da quebra espontânea de simetria, por meio da utilização de uma analogia entre o movimento de osciladores anarmônicos acoplados e a geração de massa via campo de Higgs. Complementamos a discussão feita em \cite{Helayel} acrescentando efetivamente dois aspectos: primeiro, a interação entre dois osciladores, de modo a demonstrar o mecanismo de geração de massa e, segundo, uma modelagem para a mudança de sinal da constante da mola para justificar o potencial de quebra de simetria para baixas energias. 

\noindent\hspace{1.0cm}Apresentamos também propostas de práticas de simulação computacional em sala de aula com o objetivo de solidificar as ideias descritas. Mostramos gráficos dos potenciais envolvendo as partículas $A$ e $h$ para várias situações, sendo a culminância da prática a apresentação do gráfico da energia potencial da partícula $h$ na situação de quebra de simetria, quando ele fica de fato assimétrico em relação ao eixo vertical.

\noindent\hspace{1.0cm}Acreditamos que estudantes de Ensino Médio e de graduação em Física que frequentaram aulas básicas de mecânica clássica sobre o Oscilador Harmônico têm plenas condições de compreender o mecanismo de geração de massa. A utilização de simulações computacionais, como o GeoGebra, para visualizar as transições de fase e simetrias em diferentes temperaturas, reforça a eficácia do modelo no ambiente educacional. A interatividade com as ferramentas digitais podem tornar o aprendizado mais envolvente e intuitivo, facilitando a compreensão de conceitos que, de outra forma, poderiam parecer demasiado abstratos e difíceis de compreender.   

\noindent\hspace{1.0cm}A relevância do estudo se destaca ainda mais no contexto atual da física, onde o mecanismo de Higgs ocupa um lugar central, especialmente após a confirmação experimental do bóson de Higgs em 2012. Este trabalho, ao tornar esses conceitos acessíveis, contribui para a formação curricular de estudantes e dão outra perspectiva para enfrentar os desafios e avanços da área.

\noindent\hspace{1.0cm}Os resultados obtidos ao longo deste estudo mostram que o modelo é não apenas eficaz do ponto de vista pedagógico, mas também flexível o suficiente para ser expandido para outros contextos, como a aplicação a simetrias internas contínuas, além de aspectos termodinâmicos associados. Isso sugere que o modelo pode ser uma ferramenta poderosa para introduzir estudantes a uma variedade de fenômenos físicos complexos, utilizando uma base de conhecimento já familiar a eles. Também é possível a construção de diversas sequências didáticas de ensino sobre este tema, abordando seus diferentes aspectos e comparando aprendizagens associadas.

\noindent\hspace{1.0cm}Outras simetrias podem ser construídas em modelos desse tipo, ainda mais elaborados, o que pode certamente ajudar em uma maior compreensão de aspectos mais profundos de quebras de simetria em Física de Partículas. Em particular, o trabalho \cite{Perovano:2024dnq} propõe um modelo, ainda na linha de osciladores anarmônicos, para se discutir simetrias de calibre. É nossa perspectiva fazer também uma contribuição neste sentido, com as devidas adaptações didáticas com foco no ensino em educação básica. Outra ideia que pode ser estudada está associada com Teorias de Campos zero dimensionais, que pretendemos discutir em trabalhos futuros.

 \section*{Agradecimentos} Os autores agradecem ao Conselho Nacional de Desenvolvimento Científico e Tecnológico - CNPq e à FUNCAP, via projeto BP5-0197-00117.01.00/22, pelo apoio financeiro.
%%%%%%%%%%%%%%%%%%%%%%%%%%%%%%%%%%%%%% B I B L I O G R A F I A %%%%%%%%%%%%%%%%%%%%%%%%

\end{document}